\documentclass[12pt]{article}
\usepackage{amssymb}
\usepackage{epsfig}
\usepackage{float}

\usepackage{graphicx}
\begin{document}
\begin{center}
{\bf Neutrino masses and oscillations}\footnote{ A report at the symposium
"100th anniversary of the discovery of atomic nucleus"
March 10-11, 2011, JINR, Dubna, Russia.}
\end{center}
\begin{center}
S. M. Bilenky
\end{center}

\begin{center}
{\em  Joint Institute for Nuclear Research, Dubna, R-141980,
Russia.\\}
{\em Physik-Department E15, Technische Universit\"at M\"unchen,
D-85748 Garching, Germany.\\}
{\em TRIUMF
4004 Wesbrook Mall,
Vancouver BC, V6T 2A3
Canada.\\}
\end{center}
\begin{abstract}
 A report on neutrino masses, mixing and oscillations, made in Dubna at the symposium  dedicated to 100 years of the Rutherford's discovery of atomic nucleus, is presented. We start with the hypothesis of neutrino which was proposed by W. Pauli in December  1930 in order to solve some problems of nuclei (the problem of  spin of $^{7}N_{14}$ and other nuclei and the problem of continuous $\beta$-spectra).
After that we consider the theory of massless two-component neutrino and first ideas of neutrino oscillations which were put forward by B. Pontecorvo in Dubna in 1957-58. The present status of neutrino mixing and oscillations is briefly reviewed. The seesaw mechanism of the generation of small Majorana neutrino masses is discussed  and neutrinoless double $\beta$-decay of nuclei is considered. A possibility to probe
the  Majorana mass mechanism of the $0\nu\beta\beta$-decay is discussed.
\end{abstract}

\section{Introduction}

Idea of neutrinos was proposed by W. Pauli on December 4th 1930 in order to solve a problem of spin of $^{7}\mathrm{N}_{14}$ and other nuclei and to explain a continuous $\beta$-spectra of nuclei.

During many years after the Rutherford's discovery of atomic nucleus nuclei were considered as bound states of protons and electrons, the only known at that time elementary particles. There were two fundamental problems in the framework of this assumption.
\begin{enumerate}
  \item The $\beta$-decay of a nucleus in this model is  two-particle decay $(A,Z)\to (A,Z+1)+e^{-}$. From the conservation of energy and momentum follows in this case that electrons produced in a $\beta$-decay must have definite energy, practically equal to the mass difference of the initial and final nuclei. In experiments, however, continuous $\beta$-spectra were observed.
  \item There was also a problem of spin of some nuclei. For example, $^{7}\mathrm{N}_{14}$ nucleus from the point of view of proton-electron model  is a bound state of 14 protons and 7 electrons. It must have half-integer spin. On the other side, from measurements of spectra of molecular nitrogen followed that $^{7}\mathrm{N}_{14}$ nuclei  satisfy Bose-Einstein statistics and according to the theorem on connection between spin and statistics the spin of $^{7}\mathrm{N}_{14}$ nucleus must be integer.
\end{enumerate}
Analyzing these nuclear problems Pauli came to the conclusion that they can be solved only if we assume that {\em exist a new particle.} In order to explain continuous $\beta$-spectra we need to assume that in the $\beta$-decay of a nucleus not only electron but also another particle is emitted. Pauli called this particle a "neutron". He assumed that the "neutron" has spin 1/2 and its interaction with matter is much weaker that the interaction of photon (in order not to be observed in the $\beta$-decay experiments).
Pauli also assumed the "neutron" together with protons and electrons is a constituent of nuclei. This allowed him  to solve the problem of spin of $^{7}\mathrm{N}_{14}$ and other nuclei.

Pauli assumed that the new particle has a mass. In his famous letter addressed to the participants of the nuclear conference in T\"ubingen he wrote "The mass of the neutrons should be of the same order of magnitude as the electron mass and
in any event not larger than 0.01 of the proton mass".

In 1932  soon after the discovery of the neutron by Chadwick it was suggested by Heisenberg, Majorana and Ivanenko that nuclei are bound states of protons and neutrons. The proton-neutron theory could explain all nuclear data. The problem of the spin of $^{7}\mathrm{N}_{14}$ and other  nuclei disappeared. In fact, in the framework of proton-neutron model $^{7}\mathrm{N}_{14}$  is a bound state of 7 protons and 7 neutrons and has integer spin (in accordance with experimental data). What about
$\beta$-decay and continues $\beta$-spectrum?

In the framework of proton-neutron structure of nuclei the problem of the $\beta$-decay of nuclei was solved by E. Fermi in 1933-34 \cite{Fermi}. Fermi accepted the Pauli hypothesis of the existence of a neutral, light, weakly interacting particle which is emitted together with electron in the $\beta$-decay of nuclei. After discovery of the neutron he proposed to call this particle neutrino (from Italian {\em neutral, small}). Fermi  understood that {\em electron-neutrino pair is  produced in the quantum transition of a neutron inside of a nucleus into the proton}
\begin{equation}\label{transition}
n\to p +e^{-} +\bar\nu
\end{equation}
By analogy with electrodynamics Fermi proposed the first (vector) Hamiltonian of the
$\beta$-decay
\begin{equation}\label{Hamiltonian}
\mathcal{H}_{I}=G_{F}\bar p\gamma^{\alpha}n~\bar e\gamma_{\alpha}\nu + \rm{ h.c.},
\end{equation}
where $G_{F}$ is a constant which has dimension $M^{-2}$.
The Hamiltonian (\ref{Hamiltonian}) provides the transition (\ref{transition}).

In the paper on the $\beta$-decay Fermi proposed a method of the measurement of the neutrino mass. The method was based on the high-accuracy measurement of the end-point part of the electron spectrum in which neutrino energy is comparable with neutrino mass.\footnote{The same method of the measurement of the neutrino mass was proposed also by Perrin\cite{Perrin}.}

The Fermi-Perrin idea was realized in experiments  performed by  G. Hanna and B. Pontecorvo \cite{HannaPonte} and S. Curran et al \cite{Curran} in 1949. In these experiments it was found that the upper bound of the neutrino mass is much smaller than the electron mass
\begin{equation}\label{numass}
    m_{\nu}\leq 500~\mathrm{eV}\simeq 10^{-3}~m_{e}.
\end{equation}
Idea of neutrino and neutrino mass is drastically changed after it was discovered that in weak processes the parity $P$ (and charge conjugation $C$)
are violated (1957). In order to explain large violation of the parity  in the $\beta$-decay and other weak decays Landau \cite{Landau}, Lee and Young \cite{LeeYang} and Salam
\cite{Salam} proposed the theory of {\em the two-component neutrino}.

Taking into account that the upper bound of the neutrino mass is much smaller than the electron mass, Landau, Lee and Young  and Salam assumed that neutrino mass is equal to zero. In this case the left-handed (right-handed) component of the neutrino field  $\nu_{L}(x) = \frac{1- \gamma_{5}}{2}~\nu(x)$ ($\nu_{R}(x) = \frac{1+ \gamma_{5}}{2}~\nu(x)$) satisfies the Dirac equation
\begin{equation}\label{Dirac}
i\gamma^{\alpha} \partial_\alpha\nu_{L,R}(x)=0.
\end{equation}
Thus, if neutrino mass is equal to zero, the neutrino field can be $\nu_{L}(x)$ or $\nu_{R}(x)$. For the most general four-fermion Hamiltonian of the $\beta$-decay we have in this case
\begin{equation}\label{beta}
 {\cal{H}}_{I}= \sum_{i}G_{i}\, \bar{p}~ O_{i}n~
\bar{e}~ O^{i}\frac{1}{2}(1\mp \gamma_{5})  \nu  + \rm{ h.c.}.
\end{equation}
Here
$$O_{i} \to  1,\, \gamma_{\alpha},\,\sigma_{\alpha\,\beta},\,
\gamma_{\alpha}\gamma_{5},\, \gamma_{5}.$$
and $G_{i}$ are  interaction constants.

 It follows from (\ref{beta}) that in the case of the two-component neutrino theory the scalar and pseudoscalar parts of the Hamiltonian  are characterized by the same constants. This assures large violation of parity in the $\beta$-decay in accordance with the result obtained in the Wu et al.\cite{Wu} and other experiments  .

According to the two-component neutrino theory neutrino is a particle with definite helicity (negative in the case of $\nu_{L}(x)$-field and positive in the case of $\nu_{R}(x)$-field). This important prediction of the two-component neutrino theory was perfectly confirmed in the spectacular  Goldhaber et al. experiment \cite{Goldhaber}. In this experiment the circular polarization of $\gamma$'s produced in the chain
of reactions
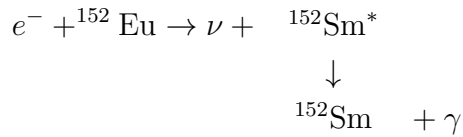
\begin{eqnarray}\label{chain}
e^- + ^{152}\rm{Eu} \to \nu + \null & ^{152}\rm{Sm}^* & \nonumber
\\
& \downarrow & \nonumber
\\
& ^{152}\rm{Sm} & \null + \gamma \nonumber
\end{eqnarray}
was measured. Spins of $^{152}\rm{Eu}$ and $^{152}\rm{Sm}$ are equal to zero and the spin of $^{152}\rm{Sm}^{*}$ is equal to one. From the conservation of momentum follows that from the measurement of the circular polarization of $\gamma$'s the neutrino helicity can be determined. It was shown in the Goldhaber at al. experiment that neutrino is a particle with negative helicity. Thus, the two-component neutrino theory was confirmed. From two possibilities ($\nu_{L}(x)$ or $\nu_{R}(x)$) the experiment allowed to choose $\nu_{L}(x)$.

The two-component neutrino theory played an extremely important role in the development of the theory of weak interaction (inspite, as we know now neutrino masses are different from zero). The phenomenological $V-A$ theory of Feynman and Gell-Mann\cite{FeyGel} and Marshak and Sudarshan \cite{MarSudar} was a generalization of the two-component neutrino theory: it was based on the assumption that in the charged current (CC) Hamiltonian of the weak interaction enter {\em left-handed components of  all fields}. The unified theory of the electroweak interaction (Standard Model) is based on the assumption that left-handed  quark, lepton and neutrino fields are components of $SU(2)$ doublets and, correspondingly, only left-handed fields enter into the CC electroweak interaction.

On the other side, because of the two-component neutrino theory {\em during many years physicists believed that neutrinos are massless particles}.

The first physicist who started to think  about a possibility of a small neutrino mass was B. Pontecorvo (1957-58). He believed in a symmetry between weak interaction of hadrons and leptons and look for an analogy of the $K^{0}\leftrightarrows \bar K^{0}$
oscillations in the lepton sector. He considered first $(\mu^{-}e^{+})\leftrightarrows
(\mu^{+}e^{-})$ oscillations. In the paper on  muonium-antimuonium oscillations \cite{Ponte1}  Pontecorvo mentioned a possibility of neutrino oscillations:

"If the two-component neutrino theory turn out to be incorrect
(which at present seems to be rather improbable)
and if the conservation law of neutrino charge would not apply,
then in principle neutrino $\rightleftarrows $ antineutrino transitions
could take place in vacuum."

At that time only one type of neutrino was known. By analogy with $K^{0}-\bar K^{0}$ system B. Pontecorvo assumed that
\begin{equation}\label{mix}
|\nu_{L}\rangle= \frac{1}{\sqrt{2}}(|\nu_{1L}\rangle+ | \nu_{2L}\rangle,\quad
|\bar \nu_{L}\rangle= \frac{1}{\sqrt{2}}(|\nu_{1L}\rangle- | \nu_{2L}\rangle),
\end{equation}
 where$|\nu_{1L}\rangle$ and $|\nu_{2L}\rangle$ are states of the Majorana neutrinos with masses $m_{1}$ and $m_{1}$ and negative helicity,
$|\nu_{L}\rangle$ is the state of the left-handed neutrino
and $|\bar \nu_{L}\rangle$ is the state of the left-handed antineutrino \footnote{According to the two-component theory with the left-handed field $\nu_{L}(x)$ neutrino is left-handed particle and antineutrino is right-handed particle. The left-handed antineutrino (right-handed neutrino) are quanta of the right-handed field $\nu_{R}(x)$ which does not enter into the Hamiltonian of the weak interaction. Thus, $\nu_{R}$ and  $\bar\nu_{L}$ have no weak interaction.
B. Pontecorvo proposed to call these particles {\em sterile}.}

In the first paper on neutrino oscillations \cite{Ponte2} B. Pontecorvo wrote:

"...the number of events $\bar \nu +p\to e^{+}+n$ with reactor antineutrino
would be smaller than the expected number. It would be
extremely interesting to perform  the Reins-Cowan experiment at different distances
from reactor." This prediction many years later was confirmed in the KamLAND reactor experiment.

In 1967 in paper \cite{Ponte3} B. Pontecorvo considered all possible transitions between $\nu_{e}$ and
$\nu_{\mu}$ and applied the idea of neutrino oscillations to the solar neutrinos. He wrote

"..due to neutrino oscillations the observed flux of solar neutrinos could be two times smaller than the expected flux"

Three years later in the Homestake neutrino experiment by R. Davis the first data were obtained (see \cite{Homestake}). It was found in this experiment that the upper bound on the flux of the solar neutrinos  was (2-3) times smaller than the flux predicted by the standard solar model. This was called the solar neutrino puzzle. B. Pontecorvo envisaged "the puzzle".

In 1969 V. Gribov and B. Pontecorvo \cite{GribovPonte} considered two-neutrino oscillations in the case of two massive Majorana neutrinos. They derived the expression for the two-neutrino transition probability in vacuum and applied it to the solar neutrinos.

Starting from  1975 many papers on neutrino oscillations were published by S. B. and
B. Pontecorvo (see \cite{BilPonte}). We developed phenomenological theory of the neutrino oscillations, consider all possible neutrino mass terms (Dirac, Majorana, Dirac and Majorana) and discussed possible neutrino oscillation experiments.

At that time nobody knew values of neutrino masses and mass-squared differences. Our main ideas were the following
\begin{enumerate}
  \item Neutrino oscillations necessary to search for in all possible neutrino experiments (solar, reactor, accelerator, atmospheric) because they are sensitive to different values of neutrino mass-squared differences.
  \item Because neutrino oscillations is an interference phenomenon the search for neutrino oscillations is the most sensitive way to look for small neutrino mass-squared differences.
\end{enumerate}
In 1977 at the time when majority of physicists believed in two-component massless neutrinos we wrote the first review on neutrino oscillations \cite{BilPonte1}. This review attracted attention of many physicists to the problem of neutrino masses, mixing and oscillations.

In four years after the first B. Pontecorvo paper on neutrino oscillations, Maki, Nakawa and Sakata \cite{MNS} on the basis of a  model, in which nucleons were considered as a bound states of neutrinos and some vector bosons, came to an idea of massive neutrinos and neutrino mixing. In connection with Brookhaven neutrino experiment (1962) they discussed "virtual transmutation" of $\nu_{\mu}\to \nu_{e}$.

In the eighties special experiments on the search for neutrino oscillations with neutrinos from reactors and accelerators were performed. No indications in favor of the neutrino oscillations were found in these experiments. On the other side, in experiments with solar and atmospheric neutrinos some evidence for neutrino oscillations was obtained. This was the situation with neutrino oscillations before 1998.

{\em In 1998 golden years of neutrino oscillations started.} In 1998 the first model independent evidence for neutrino oscillations was obtained in the Super-Kamiokande experiment \cite{SK}. In this experiment a significant up-down asymmetry of the atmospheric high-energy muon events was discovered. In 2001-2004 in the SNO solar neutrino experiment model independent proof of the transition of solar  $\nu_{e}$ into  $\nu_{\mu}$ and  $\nu_{\tau}$ was obtained \cite{SNO}. In this experiment it was shown that the ratio of the flux of the solar $\nu_{e}$ to the total flux of $\nu_{e}$, $\nu_{\mu}$ and $\nu_{\tau}$ is approximately equal to 1/3. In 2002-2004 in the reactor KamLAND experiment \cite{Kamland} a model independent evidence in favor of the disappearance of the reactors $\bar\nu_{e}$ was found. In this experiment a significant distortion of the spectrum of the reactor antineutrinos was found.

\section{Neutrino oscillations}

We start with the  discussion of {\em the basics of the neutrino oscillation}. Analysis of the neutrino oscillations is based on the following assumptions (see, for example, \cite{BGG})
\begin{enumerate}
  \item Interactions of flavor neutrinos is given by the Standard Model CC and NC Lagrangians.  The Lagrangian of the CC interaction of neutrinos, leptons and $W$ is given by the expression
\begin{equation}\label{CCinteraction}
\mathcal{L^{CC}_{I}}(x)=-\frac{g}{2\sqrt{2}}\sum_{l=e,\mu,\tau}\bar \nu_{l L}(x) \,\gamma_{\alpha}\, l_{L}(x)~W^{\alpha}(x)+
\mathrm{h.c.}
\end{equation}
All existing weak interaction data are in agreement with this assumption.
 \item The field of the flavor neutrinos $\nu_{l }$ is the  "mixed field"
\begin{equation}\label{mixfield}
\nu_{l L}(x)=\sum^{3}_{i=1} U_{l i}\,\nu_{i L}(x).
\end{equation}
Here $U$ is an unitary $3\times 3$ PMNS \cite{Ponte1,MNS} mixing matrix and
 $\nu_{i }(x)$ is the field of neutrino (Dirac or Majorana) with mass $m_{i}$.
\item
From (\ref{mixfield})  follows  that flavor neutrino $\nu_{l}$, which is produced in   CC weak processes together with $l^{+}$, is described by the mixed state
\begin{equation}\label{mixstate}
|\nu_{l}\rangle=\sum^{3}_{i=1} U^{*}_{l i}\,|\nu_{i}\rangle,
\end{equation}
where $|\nu_{i}\rangle$ is the state of neutrino with mass $m_{i}$ and 4-momentum $p_{i}$.
\item The probability of the transition $\nu_{l} \to\nu_{l'}$
in vacuum is given by the following expression
\begin{equation}\label{Transition}
\mathrm{P}(\nu_{l} \to\nu_{l'}) =|\sum^{3}_{i=1} U_{l'i} \,e^{-i\frac{\Delta m^2_{ki }L }{ 2E}}\,U^{*}_{l i}|^{2}= |\delta_{l'l}+
\sum_{i\neq k} U_{l'i} \,(e^{-i\frac{\Delta m^2_{ki }L }{ 2
E}}-1)\,U^{*}_{l i}|^{2}.
\end{equation}
Here $\Delta m^2_{ki }= m^2_{i }-m^2_{k }$, $L$ is the distance between  neutrino source and  neutrino detector and $E$ is the neutrino energy.
\end{enumerate}
The expression (\ref{Transition}) follows from the Schrodinger equation for the quantum states. In fact, we have
\begin{equation}\label{evolution}
|\nu_{l}\rangle_{t}=e^{-iHt}~|\nu_{l}\rangle=\sum^{3}_{i=1}e^{-iE_{i}t}  U^{*}_{l i}\,|\nu_{i}\rangle=\sum_{l'=e,\mu,\tau}|\nu_{l'}\rangle(\sum^{3}_{i=1}U_{l'i}e^{-iE_{i}t}  U^{*}_{l i}).
\end{equation}
For ultrarelativistic neutrinos we have
\begin{equation}\label{evolution1}
E_{i}\simeq p_{i}+\frac{m^{2}_{i}}{2E}
\end{equation}
Assuming $p_{i}=p_{k}$, from (\ref{evolution}) and (\ref{evolution1}) we easily obtain the expression (\ref{Transition}).

From the data of the LEP experiments on the measurement of the width of the decay $Z\to \nu_{l}+\bar\nu_{l}$ follows that the number of the flavor neutrinos is equal to three ($n_{\nu}=2.9840\pm 0.0082 $).

If the number of massive neutrinos $\nu_{i}$ is equal to the number of the flavor neutrinos,  neutrino transition probabilities in vacuum
are characterized by six parameters: two mass-squared differences, three mixing angles $\theta_{12}$, $\theta_{23}$ and $\theta_{13}$
and $CP$ phase $\delta$.

Two neutrino mass spectra are compatible with existing data in the case of the tree massive neutrinos:
\begin{enumerate}
\item  Normal spectrum (NS)
 $m_{1}<m_{2}<m_{3}, \quad \Delta m^{2}_{12}\ll\Delta m^{2}_{23}$
 \item Inverted spectrum (IS)  $m_{3}<m_{1}<m_{2}, \quad \Delta m^{2}_{12}\ll|\Delta  m^{2}_{13}|$
\end{enumerate}
Notices that notations for neutrino masses are different for NS and IS spectra. Let us determine atmospheric and solar mass-squared differences in the following way:
\begin{equation}\label{notations}
\Delta m^2_{A }=\Delta m^2_{23 }(\mathrm{NS})=|\Delta  m^{2}_{13}| (\mathrm{IS}),~~
\Delta m^2_{S }=\Delta m^2_{12 }(\mathrm{NS})=\Delta  m^{2}_{12} (\mathrm{IS}).
\end{equation}
From analysis of the existing neutrino oscillation data follows that $\Delta m^2_{S }$ is much smaller than $\Delta m^2_{A }$ and  the angle
$\theta_{13}$ is small:
\begin{equation}\label{smallness}
\Delta m^2_{S }\simeq \frac{1}{30}\Delta m^2_{A }, \quad \sin^{2}\theta_{13}\lesssim 4\cdot 10^{-2}.
\end{equation}
If we take into account these relations we can easily show (see \cite{BGG}) that in atmospheric, reactor and accelerator neutrino oscillation experiments dominant transitions are described by simple two-neutrino expressions. In fact, in atmospheric and accelerator experiments we have
$\frac{\Delta m^2_{A }L }{2E}\simeq 1$ and $\frac{\Delta m^2_{S }L }{2E}\ll 1$.
Neglecting small contributions of $\frac{\Delta m^2_{S }L }{2E}$ and $\sin^{2}\theta_{13}$, we easily find that dominant transitions in atmospheric and accelerator experiments are
$\nu_{\mu} \to\nu_{\tau}$ and $\bar\nu_{\mu} \to\bar\nu_{\tau}$. For the $\nu_{\mu}$ ($\bar\nu_{\mu}$) survival probability from (\ref{Transition}) we find the following expression
\begin{equation}\label{2atmosph}
\mathrm{P}(\nu_{\mu}\to\nu_{\mu})=\mathrm{P}(\bar\nu_{\mu}\to\bar\nu_{\mu})
\simeq 1-\frac{1}{2}\sin^{2}2\theta_{23}~
(1-\cos\Delta m_{A}^{2}\frac{L}{2E}).
\end{equation}
Neglecting contribution of $\sin^{2}\theta_{13}$, for the $\bar\nu_{e}$ survival probability
in reactor KamLAND experiment ($\frac{\Delta m^2_{S }L }{2E}\simeq 1$) we have
\begin{equation}\label{2react}
\mathrm{P}(\bar\nu_{e}\to\bar\nu_{e})=1-\sum_{l=\mu,\tau} \mathrm{P}(\bar\nu_{e}\to\bar\nu_{l})
\simeq 1-\frac{1}{2}\sin^{2}2\theta_{12}~
(1-\cos\Delta m_{S}^{2}\frac{L}{2E}).
\end{equation}
Thus, in the atmospheric region of the parameter $\frac{L}{E}$ the leading oscillations
($\nu_{\mu} \leftrightarrows\nu_{\tau}$ and $\bar\nu_{\mu} \leftrightarrows \bar\nu_{\tau}$)
are determined by the parameters $\Delta m_{A}^{2}$ and $\sin^{2}2\theta_{23}$. In the reactor KamLAND region of $\frac{L}{E}$ the leading oscillations
($\bar\nu_{e} \leftrightarrows\bar\nu_{\mu,\tau}$) are determined by the parameters  $\Delta m_{S}^{2}$ and $\sin^{2}2\theta_{12}$. Let us notice also that in the leading approximation the probability of solar $\nu_{e}$ to survive  is given by the two-neutrino expression for neutrino transition in matter which depend on the parameters $\Delta m_{S}^{2}$ and $\sin^{2}2\theta_{12}$ and the number density of electrons in the sun. The existing neutrino oscillation data are well described by the leading approximation.

From the two-neutrino analysis of the  data of the long baseline accelerator experiment MINOS \cite{Minos} it was inferred
\begin{equation}\label{Minos}
\Delta m_{A}^{2}=2.32^{+0.12}_{-0.08}\cdot 10^{-3}~\mathrm{eV}^{2},\quad \sin^{2}2\theta_{23}>0.90~ (90\% CL).
\end{equation}
These values of the oscillation parameters $\Delta m_{A}^{2}$ and $\sin^{2}2\theta_{23}$ are in a good agreement with the values of the parameters obtained from the three-neutrino analysis of the data of the atmospheric Super-Kamiokande experiment\cite{SK}. For  the normal (inverted) mass spectrum it was found
\begin{equation}\label{SKdata}
 1.9~ (1.7)\cdot 10^{-3}  < \Delta m_{A}^{2}<2.6~ (2.7)\cdot 10^{-3} ~\mathrm{eV}^{2},~~~0.407<\sin^{2}\theta_{23}<0.583
\end{equation}
For the parameter $\sin^{2}\theta_{13}$ the following upper bound was obtained
\begin{equation}\label{SKdata1}
 \sin^{2}\theta_{13}<0.04 ~(0.09).
\end{equation}
From three-neutrino analysis of the solar and reactor KamLAND data \cite{Kamland} it was found
\begin{equation}\label{kamland}
\Delta m_{S}^{2}=7.50^{+0.19}_{-0.20}\cdot 10^{-5}~\mathrm{eV}^{2},~~ \tan^{2}\theta_{12}=0.452^{+0.035}_{-0.033},
~~\sin^{2}\theta_{13}=0.020^{+0.016}_{-0.016}.
\end{equation}
From analysis of the reactor CHOOZ data \cite{Chooz} the following bound was obtained
\begin{equation}\label{chooz}
\sin^{2}\theta_{13}\leq 4\cdot 10^{-2}.
\end{equation}
Summarizing, from the analysis of the  data of the present-day neutrino oscillation experiments four neutrino oscillation parameters $\Delta m_{S}^{2}$, $\Delta m_{A}^{2}$,
$\tan^{2}\theta_{12}$ and $\sin^{2}2\theta_{23}$ were determined with accuracies (3-10)\%. For the parameter $\sin^{2}\theta_{13}$ only upper bound was inferred from existing data. There is no information about the value of the $CP$ phase $\delta$.

In future accelerator (T2K \cite{T2K}, NOvA \cite{Nova}) and reactor (Double Chooz\cite{Doublechooz}, RENO\cite{Reno}, Daya Bay\cite{Dayabay}) experiments sensitivity to the parameter  $\sin^{2}\theta_{13}$ will be one order of magnitude higher than in the CHOOZ experiment.

If the value of the parameter $\sin^{2}\theta_{13}$ is relatively large,  a possibility to study $CP$ violation in the lepton sector and a possibility to reveal the character of the  neutrino mass spectrum will appear.

Let us also notice that from tritium experiments (Troitsk\cite{Troitsk}, Mainz\cite{Mainz}) the following bound on  the absolute value of the neutrino mass was obtained
\begin{equation}\label{Tritium}
m_{\beta}<2.3 ~\mathrm{eV}.
\end{equation}
The sensitivity of the future tritium experiment KATRIN\cite{Katrin} will be an order of magnitude higher ($m_{\beta}\simeq 0.2 ~\mathrm{eV}$). From existing cosmological data the following bound on the sum of the neutrino masses
\begin{equation}\label{cosmobound}
    \sum_{i}m_{i}<(0.6-1.0)~\mathrm{eV}
\end{equation}
can be inferred(see \cite{Hannestad}).

Atmospheric, solar, reactor and long baseline accelerator neutrino oscillation data are well described  if we assume three-neutrino mixing.
Exist, however, data of  short baseline neutrino oscillation experiments which (if correct)
require  $3+n_{s}$ ($n_{s}=1, 2...$) massive neutrinos and, correspondingly, $n_{s}$ sterile neutrinos.

The general neutrino mixing has the form
\begin{equation}\label{sterile}
\nu_{lL}=\sum_{i=1}^{3+n_{s}}U_{li}\nu_{iL},\quad
\nu_{sL}=\sum_{i=1}^{3+n_{s}}U_{si}\nu_{iL}.
\end{equation}
Here $\nu_{i}$ is the field of neutrino with mass $m_{i}$, $U$ is an unitary
$(3+n_{s})\times (3+n_{s})$ mixing matrix and index $s$ takes the values $s_{1}, s_{2},...s_{n_{s}}$.

During many years exists an indication in favor of $\bar\nu_{\mu}\leftrightarrows\bar\nu_{e}$ oscillations obtained in the accelerator short baseline experiment  LSND\cite{LSND}. The data of this experiment can be described by the two-neutrino transition probability
\begin{equation}\label{Lsnd}
\mathrm{P}(\bar\nu_{\mu}\to\bar\nu_{e})
\simeq 1-\frac{1}{2}\sin^{2}2\theta_{e\mu}~
(1-\cos\Delta m_{14}^{2}\frac{L}{2E}),
\end{equation}
where $\sin^{2}2\theta_{e\mu}=4|U_{e4}|^{2}|U_{\mu4}|^{2}$.

From the analysis of the LSND data it was found that $\mathrm{\overline P}(\bar\nu_{\mu}\to\bar\nu_{e})=0.267\pm 0.067\pm 0.045$. For the mass-squared difference $\Delta m_{14}^{2}$ the following  range
$$0.2<\Delta m_{14}^{2}<2~\mathrm{eV}^{2}$$ was obtained.

Thus, in the LSND experiment an indications in favor of neutrino oscillations with mass-squared difference, which is much larger than solar and   atmospheric mass-squared differences,
was obtained. Confirmation of the LSND data would be an evidence in favor of the existence of four (or more) massive neutrinos and sterile neutrino(s).

In the accelerator MiniBooNE experiment\cite{Miniboone}, which was aimed to check the LSND result, no indications in favor of $\nu_ {\mu}\to\nu_{e}$ oscillations were found in the LSND $L/E$ region. However, some (about 2 $\sigma$) indications in favor of neutrino oscillations, compatible with the LSND result, were obtained in the
$\bar\nu_{\mu}\to\bar\nu_{e}$ channel \cite{Miniboone1}.

Recent recalculations of the fluxes of the reactor antineutrinos\cite{reactor1}
allow to reinterpret results of old reactor neutrino  experiments as an indication in favor of short baseline neutrino oscillations driven by
$\Delta m^{2}\simeq 1~ \mathrm{eV}^{2}$\cite{reactor2}. In order to clarify the situation with short baseline neutrino oscillations, which require mixing of (at least) four massive neutrinos, new reactor  and accelerator neutrino experiments were proposed (see \cite{Rubbia}).

\section{On the seesaw mechanism of the neutrino mass generation}

From neutrino oscillation experiments which allow to determine neutrino mass-squared differences we can conclude that {\em neutrino masses are different from zero}. On the other side, from upper bounds which were
obtained from the results of the tritium $\beta$-decay experiments and from analysis of the cosmological data it follows that {\em neutrino masses are small, much smaller than masses of quarks and leptons}.

Let us compare masses of particles of the third family. We have
\begin{equation}\label{masses}
m_{t}\simeq 1.7\cdot 10^{2}~\mathrm{GeV},~~m_{b}\simeq 4.7~\mathrm{GeV},~~
{\bf m_{3}< 2.3\cdot 10^{-9}~\mathrm{GeV}},~~m_{\tau}\simeq 1.8~\mathrm{GeV}.
\end{equation}
We believe that masses of leptons and quarks are generated by the standard Higgs mechanism. From (\ref{masses}) it is evident that {\em it is very unlikely that neutrino masses are of the same Higgs origin}. Some new (or additional) mechanism of the neutrino mass generation is necessary.

Several such mechanisms were proposed. The most plausible and viable is the seesaw mechanism of the neutrino mass generation\cite{seesaw}. This mechanism is based on the assumption that neutrino masses are generated due to a beyond the SM physics at a scale $\Lambda$ which is much larger than the electroweak scale $v=(\sqrt{2}G_{F})^{-1}\simeq 246$ GeV ($v$ is the Higgs vacuum expectation value).

At relatively small electroweak energies a new beyond the SM physics induce non-renormalizable effective Lagrangian of the form
\begin{equation}\label{effectiveL}
    \mathcal{L}^{\mathrm{eff}}=\sum_{n\geq 1}\frac{1}{\Lambda^{n}}\mathcal{L}_{4+n},
\end{equation}
where the term $\mathcal{L}_{4+n}$, which is built from the SM fields, is invariant under $SU_{L}(2)\times U(1)$ transformations and has dimension $M^{4+n}$.

The largest contribution comes from dimension 5 term of the effective Lagrangian.
{\em The only dimension 5 term  has the form} \cite{Weinberg}
\begin{equation}\label{nueffective}
\mathcal{L}_{5}^{\mathrm{eff}}=-\frac{1}{\Lambda}
\sum_{l',l}\,\overline L_{l'L}\tilde{H }Y_{l'l}C
\tilde{H }^{T}(\overline L_{lL})^{T}
+
\mathrm{h.c.},
\end{equation}
where
\begin{eqnarray}\label{heavyH1}
L_{lL}
=
\left(
\begin{array}{c}
\nu_{lL}
\\
l_L
\end{array}
\right),
\qquad
H
=
\left(
\begin{array}{c}
H^{(+)}
\\
H^{(0)}
\end{array}
\right)
\end{eqnarray}
are lepton and Higgs doublets, $\tilde{H}=i\tau_{2}H^{*}$ , $C$ is the matrix of the charge conjugation and $Y$ is a complex symmetrical matrix. It is important to stress that the Lagrangian (\ref{nueffective}) does not conserve the total lepton number $L=L_{e}+L_{\mu}+L_{\tau}$.\footnote{The effective Lagrangian (\ref{nueffective}) can be generated in the second order of the perturbation theory by the lepton number violating interaction of a heavy  singlet (triplet) Majorana fermions with lepton and Higgs doublets (type I(III) seesaw) and by the $L$-violating interaction
of  lepton doublets and a Higgs doublet with a  heavy triplet scalar boson
(type II seesaw).}

If we put
\begin{eqnarray}\label{heavyH2}
\tilde{H}
=
\left(
\begin{array}{c}
\frac{v}{\sqrt{2}}
\\
0
\end{array}
\right),
\end{eqnarray}
the electroweak symmetry will be spontaneously broken and
the Lagrangian (\ref{heavyH2})
generates {\em the left-handed Majorana  mass term}
\begin{equation}\label{Mjmass}
\mathcal{L}^{\mathrm{M}}=-
\frac{1}{2}
\,\sum_{l'l}
\overline \nu_{l'L}~
M^{L}_{l'l}~C
\bar\nu_{lL}^{T}
+
\mathrm{h.c.},
\end{equation}
where
\begin{equation}\label{Mjmass1}
M^{L}_{l'l}=\frac{v^{2}}{\Lambda}Y_{l'l}.
\end{equation}
After the standard procedure of the diagonalization of the symmetrical matrix $Y$ (see \cite{BilPet}) we have
\begin{equation}\label{diagon}
Y=UyU^{T},
\end{equation}
where $U$ is a unitary matrix, $y_{ik}=y_{i}\delta_{ik},\quad y_{i}>0$

From (\ref{Mjmass}) and (\ref{Mjmass1}) we find
\begin{equation}\label{Mjmass2}
\mathcal{L}^{\mathrm{M}}=-\frac{1}{2}
\,\sum^{3}_{i=1}m_{i}\bar \nu_{i} \nu_{i},
\end{equation}
where
\begin{equation}\label{Mjmass3}
m_{i}=\frac{v^{2}}{\Lambda}y_{i}
\end{equation}
and $\nu_{i}(x)$ is the field of neutrino with mass $m_{i}$.

The field $\nu_{i}(x)$  satisfies the condition
\begin{equation}\label{Majorana}
 \nu^{c}_{i}(x)=C\bar\nu^{T}_{i}(x)=\nu_{i}(x).
\end{equation}
Thus, $\nu_{i}(x)$ is the field of the Majorana neutrino  with mass $m_{i}$. The flavor field $\nu_{lL}(x)$ is connected with the left-handed components $\nu_{iL}(x)$ by the mixing relation
\begin{equation}\label{mixing}
\nu_{lL}(x)=\sum^{3}_{i=1}U_{li}\nu_{iL}(x).
\end{equation}
From (\ref{Mjmass3}) follows that values of the neutrino masses are determined by the seesaw factor
\begin{equation}\label{seesaw}
\frac{v^{2}}{\Lambda}\simeq \frac{\mathrm{(EWscale)^{2}}}{\mathrm{scale~ of~ a~ new~ physics}}.
\end{equation}
From existing data we can estimate that $\Lambda \simeq (10^{14}-10^{15})$~ GeV.

Thus, small Majorana neutrino masses and neutrino mixing are the only observable signature of a new lepton number violating physics at a large (GUT) scale.

How can we test this idea? First of all we need to prove that neutrinos with definite masses $\nu_{i}$ are Majorana particles. We can not prove this by the investigation of neutrino oscillations\cite{BilHos}. In fact, elements of mixing matrices in the Dirac and Majorana cases differ only by  phases:
\begin{equation}\label{DMmatrices}
 U_{li}^{\mathrm{M}} =U_{li}^{\mathrm{D}}e^{i\alpha_{i}}.
\end{equation}
 It is obvious that Majorana phases $\alpha_{i}$ do not enter into transition probability (\ref{Transition}).

In order to reveal the Majorana nature of neutrinos with definite masses we need to study processes in which the total lepton number $L$ is violated. The highest sensitivity to small Majorana neutrino mass can be reached via the investigation of the neutrinoless double $\beta$-decay ($0\nu\beta\beta$-decay) \begin{equation}\label{2beta}
(A,Z)\to (A,Z+2)+e^{-}+e^{-}
\end{equation}
of $^{76}\mathrm{Ge}$, $^{130}\mathrm{Te}$, $^{136}\mathrm{Xe}$ and other even-even nuclei.
\section{Neutrinoless double $\beta$-decay}
The neutrinoless double $\beta$-decay is the second order in the Fermi constant $G_{F}$ process with  virtual neutrinos. The propagator of the mixed fields $\nu_{eL}(x)$ is given by the expression (see\cite{bbBil}
\begin{eqnarray}\label{propagator}
&&\langle 0|T(\nu_{eL}(x_{1})\nu^{T}_{eL}(x_{2}))|0\rangle=
-\frac{i}{(2\pi)^{4}}\sum_{i} U^{2}_{ei}\int d^{4}p e^{-ip(x_{1}-x_{2})}\left (\frac{1-\gamma_{5}}{2}\right )\frac{\gamma\cdot p+m_{i}}{p^{2}-m^{2}_{i}} \nonumber\\
&&\times\left (\frac{1-\gamma_{5}}{2}\right )C\simeq - m_{\beta\beta}\frac{i}{(2\pi)^{4}}\int e^{-ip(x_{1}-x_{2})}\frac{1}{p^{2}}d^{4}p
\left (\frac{1-\gamma_{5}}{2}\right )C,
\end{eqnarray}
where
\begin{equation}\label{Mjmass4}
m_{\beta\beta}=\sum_{i} U^{2}_{ei}m_{i}
\end{equation}
is the effective Majorana mass. The half-life of the $0\nu\beta\beta$-decay is given by the following general expression\cite{Doi}
\begin{equation}\label{halflife}
\frac{1}{T_{1/2}(A,Z)}=
|m_{\beta\beta}|^{2}\,|M(A,Z)|^{2}\,G(E_{0},Z).
\end{equation}
Here $M(A,Z) $ is the nuclear matrix element (NME) and  $G^{0\,\nu}(E_{0},Z)$
is a known phase-space factor ($E_{0}$ is the energy release).

No evidence for the $0\nu\beta\beta$-decay was found up to now.\footnote{Indication in favor of $0\nu\beta\beta$-decay of $^{76}\mathrm{Ge}$ which was claimed by the some participants of the Heidelberg-Moscow experiment \cite{Klapdor} will be checked in the running GERDA
experiment\cite{Gerda}.} Taking into account results of the different calculations of NME, from the data of the most precise experiments on search for the $0\mu\beta\beta$-decay of $^{76}\rm{Ge}$ \cite{baudis},
$^{130}\rm{Te}$ \cite{cuore} and $^{100}\rm{Mo}$ \cite{nemo}
the following upper bounds for the effective Majorana mass can be inferred
\begin{eqnarray}\label{bounds}
|m_{\beta\beta}| & < & (0.20-0.32) ~eV, \nonumber\\
               & < & (0.30-0.71) ~eV, \nonumber\\
               & < & (0.50-0.96) ~eV.
\end{eqnarray}
Future experiments now under preparation (CUORE\cite{Cuore}, EXO\cite{Exo}, MAJORANA
\cite{Majorana} and others) will be sensitive to\footnote{High sensitivity of $0\nu\beta\beta$-decay experiments to the Majorana neutrino mass is due to such practical reasons as large mass of a source (about 1 ton or more in future experiments),  possibilities to use low background underground laboratories, high energy resolution of $^{76}\mathrm{Ge}$ and other detectors  etc. It is interesting that most important problem of the modern neutrino physics, the problem of the nature of neutrinos with definite mass, can be solved via investigation of nuclear processes.}
\begin{equation}\label{sensitiv}
|m_{\beta\beta}|\simeq \mathrm{a~few}~10^{-2}~\mathrm{eV}.
\end{equation}
If the value of the lightest neutrino mass is smaller than $\sqrt{\Delta m_{A}^{2}}\simeq 5\cdot 10^{-2}$ eV, the value of the effective Majorana mass strongly depends on the neutrino mass spectrum. In the case of the neutrino mass hierarchy $m_{1}\ll m_{2}\ll m_{3}$ for the neutrino masses we have
\begin{equation}\label{hierarchy}
m_{2}\simeq \sqrt{\Delta m_{S}^{2}},\quad m_{3}\simeq \sqrt{\Delta m_{A}^{2}}.
\end{equation}
Neglecting the contribution of  $m_{1}$, for the effective Majorana mass we find the following upper bound
\begin{equation}\label{hierarchy1}
|m_{\beta\beta}|\simeq \left |\cos^{2}\theta_{13}\sin^{2} \theta_{12}\, \sqrt{\Delta
m^{2}_{S}}+\sin^{2} \theta_{13}\,e^{2i\alpha_{23}} \sqrt{\Delta m^{2}_{A}}\right |\lesssim 3.1 \cdot 10^{-3}~\rm{eV}.
\end{equation}
This bound is significantly smaller that the expected sensitivity of the future experiments on the search for the $0\nu\beta\beta$-decay.

In the case of the inverted mass hierarchy $m_{3}\ll m_{1}< m_{3}$ for the neutrino masses we have
\begin{equation}\label{inverted}
m_{3}\ll \sqrt{\Delta m_{A}^{2}},\quad m_{1}\simeq \sqrt{\Delta m_{A}^{2}},\quad m_{2}\simeq \sqrt{\Delta m_{A}^{2}}(1+\frac{\Delta m_{S}^{2}}{2\Delta m_{A}^{2}})\simeq\sqrt{\Delta m_{A}^{2}}.
\end{equation}
Neglecting the contribution of $\sin^{2}\theta_{13}$, for the effective Majorana mass we find
the following expression
\begin{equation}\label{inverted1}
|m_{\beta\beta}|\simeq \sqrt{ \Delta m^{2}_{A}}\,~ (1-\sin^{2}
2\,\theta_{12}\,\sin^{2}\alpha_{12})^{\frac{1}{2}},
\end{equation}
where $\alpha_{12}=\alpha_{2}-\alpha_{1}$ is the difference of the Majorana phases of the elements $U_{e2}$ and  $U_{e1}$. The parameter $\sin^{2}\alpha_{12}$ is the only unknown parameter in the expression for
$|m_{\beta\beta}|$ in the case of the inverted mass hierarchy. From (\ref{inverted1}) for e3ffective Majorana mass we have the following range
\begin{equation}\label{inverted2}
 \Delta m^{2}_{A}(1-\sin^{2}
2\,\theta_{12})    \leq |m_{\beta\beta}|^{2}\leq  \Delta m^{2}_{A},
\end{equation}
where upper and lower bounds correspond to the case of the $CP$ invariance in the lepton sector (the upper(lower) bound corresponds to the case of the equal (opposite) Majorana $CP$ parities of $\nu_{1}$ and $\nu_{2}$). Using numerical values for the atmospheric neutrino mass-squared difference and the parameter $\sin^{2}
2\,\theta_{12}$ we have
\begin{equation}\label{inverted3}
 1.7\cdot 10^{-2}    \leq |m_{\beta\beta}|\leq 5.1\cdot 10^{-2}~\mathrm{eV}
\end{equation}
Thus,
in the case of the inverted neutrino mass hierarchy the effective Majorana mass takes a value in the region of the sensitivity of the future experiments on the search for
$0\nu\beta\beta$-decay. If in these experiments the $0\nu\beta\beta$-decay will be observed it will be a proof that neutrinos with definite masses are Majorana particles and an argument in favor of the seesaw mechanism of the neutrino mass generation. {\em The measurement of the half-life of the process} would allow to probe the standard seesaw mechanism which is based on the assumption that the total lepton number is violated at a large ($\Lambda \simeq (10^{14}- 10^{15})$~ GeV scale.

In fact, from (\ref{halflife}) and (\ref{inverted2}) we obtain the following inequality\cite{BilSimPotz}
\begin{equation}\label{final}
T^{\mathrm{min}}_{1/2}(A,Z)\leq     T_{1/2}(A,Z)\leq T^{\mathrm{max}}_{1/2}(A,Z),
\end{equation}
where
\begin{eqnarray}\label{final1}
T^{\mathrm{min}}_{1/2}(A,Z)&=&\Delta m^{2}_{A}\cos^{2}
2\,\theta_{12}|M(A,Z)|^{2}G(E_{0},Z),\nonumber\\
T^{\mathrm{max}}_{1/2}(A,Z)&=&\Delta m^{2}_{A}|M(A,Z)|^{2}G(E_{0},Z)
\end{eqnarray}
If the measured half-life of the $0\nu\beta\beta$-decay of any even-even nucleus is out of the range (\ref{final}) we can conclude that the Majorana neutrino mass mechanism is not the only mechanism of the $0\nu\beta\beta$-decay. On the other side, if it will occur that the measured half-life is in the range (\ref{final}) we can not make any conclusions about the Majorana mass mechanism because of uncertainty connected with Majorana phase difference. It is evident that for that comparison we need to use the information about NME (see \cite{BilSimPotz}).

\section{Conclusion}
Neutrinos are exceptional particles. This is determined by the fact that neutrinos are the only fundamental fermions with electric charges equal to zero. In the region of energies in
 which $Q^{2}\ll m^{2}_{W}(m^{2}_{Z}$) ($Q$ is the momentum of virtual $W$ ($Z$) boson) neutrinos have only weak interaction which is characterized by the Fermi constant $G_{F}$. As a result, neutrino is an unique tool for the study of the quark structure of a nucleon (via investigation deep inelastic neutrino-nucleon scattering), for the investigation of the internal region of the sun where the sun energy is produced (via the detection of the solar neutrinos) for the study of  the mechanism of the gravitational collapse (via the detection of the supernova neutrinos) etc.

 In the last years new aspects of neutrino physics emerged. Neutrino can be an unique tool which allow to reveal a new beyond the SM physics at a very large scale  where the total lepton number is violated.

This possibility appeared after discovery in atmospheric, solar, reactor and accelerated neutrino experiments of a new phenomenon-neutrino oscillations, phenomenon which was discussed for the first time in Dubna by B. Pontecorvo as early as 1957-58.

Small neutrino mass-squared differences determined from the results of the neutrino oscillation experiments together with upper bounds on neutrino masses which were inferred from tritium $\beta$-decay experiments and cosmological data mean that neutrinos $\nu_{i}$ have masses different from zero but many orders of magnitude smaller than masses of quarks and leptons. Masses of quarks, leptons and neutrinos can not be of the same SM origin. Neutrino masses are apparently generated by a beyond the SM mechanism.

 At the electroweak scale beyond the SM physics generate nonrenormalizable effective Lagrangians.  The only dimension five term of the effective Lagrangian does not conserve the total lepton number $L$ and after spontaneous violation of the electroweak symmetry gives the Majorana neutrino mass term and seesaw suppressed Majorana neutrino masses. From this point of view the search for effects of the nonconservation of $L$, induced by small Majorana neutrino masses, is the search for a new physics at a large GUT scale. The search for neutrinoless double $\beta$-decay of even-even nuclei is the only practical way to study extremely small effects of the violation of the total lepton number due to Majorana neutrino masses. New experiments on the search for $0\nu\beta\beta$-decay sensitive to the value of effective Majorana mass
at $10^{-2}$ eV scale are in preparation and their results are eagerly waited.

\textbf{Acknowledgments}\\
It a pleasure for me to thank  Yu. Oganessian for the invitation to the  conference dedicated to 100 years of the Rutherford discovery of atomic nucleus and W. Potzel for  interesting discussions. This work was supported by funds of the Deutsche For\-schungsgemeinschaft DFG (Transregio 27: Neutrinos and Beyond), the Munich Cluster of Excellence (Origin and Structure of the Universe), and the  Maier-Leibnitz-Laboratorium (Garching).

\end{document}